\documentclass{jfm}
\usepackage{graphicx}
\usepackage{epstopdf, epsfig}
\usepackage{color}

\usepackage{amsfonts}          
\usepackage{amssymb}           
\usepackage{amsmath}   
\usepackage{multiJFMequations}
\newcommand*{\Ray}{{\rm Ra}}
\newcommand*{\Ek}{{\rm E}}
\newcommand{\Pra}{{\rm Pr}}
\newcommand{\Ro}{{\rm Ro}}
\newcommand{\Ree}{{\rm Re}}
\newcommand{\uu}{{\rm u}}

\renewcommand{\epsilon}{\varepsilon}
\newcommand{\T}{{\rm T}}
\newcommand{\dd}{{\rm d}}
\newcommand{\bigD}{{\rm D}}

\def\bfu{\mbox{\bf u}}
\def\bfe{\mbox{\bf e}}
\def\bfnabla{\mbox{\boldmath $\nabla$}}
\def\bfomega{\mbox{\boldmath $\omega$}}
\def\bfOmega{\mbox{\boldmath $\Omega$}}
\def\bfg{\mbox{\bf g}}

\def\stokes{\mbox{$\nabla_{\star}^{2}$}}

\shorttitle{Eye Formation in Rotating Convection}
\shortauthor{L. Oruba, P.A. Davidson and E. Dormy}

\title{Eye Formation in Rotating Convection}

\author{L. Oruba \aff{1},
 P.~A. Davidson \aff{2},
 \and E. Dormy \aff{3}\corresp{\email{ludivine.oruba@ens.fr, pad3@cam.ac.uk, emmanuel.dormy@ens.fr}
}}

\affiliation{\aff{1} Physics Department, Ecole Normale
  Sup\'{e}rieure,\\ 24 rue Lhomond, 75005 Paris, France.
\aff{2} Engineering Department, University of Cambridge,\\ Trumpington Street, Cambridge CB2 1PZ, UK.
\aff{3} Department of Mathematics \& Applications, CNRS UMR 8553, 
Ecole Normale Sup\'{e}rieure,\\ 45 rue d'Ulm, 75005 Paris, France.}

\begin{document}

\maketitle

\begin{abstract}
We consider rotating convection in a shallow, cylindrical domain. 
We examine the conditions under which the resulting 
vortex develops an eye at its core; that is, a region where the poloidal 
flow reverses and the angular momentum is low. For simplicity, 
we restrict ourselves to steady, axisymmetric flows in a Boussinesq 
fluid. Our numerical experiments show that, in such systems, an eye forms as 
a passive response to the development of a so-called eyewall, a conical 
annulus of intense, negative azimuthal vorticity that can form near the axis 
and separates the eye from the primary vortex. We also observe that the 
vorticity in the eyewall comes from the lower boundary layer, and relies 
on the fact the poloidal flow strips negative vorticity out of the boundary 
layer and carries it up into the fluid above as it turns upward near the axis. 
This process is effective only if the Reynolds number is sufficiently 
high for the advection of vorticity to dominate over diffusion.
Finally we observe that, in the vicinity of the eye and the eyewall, the
buoyancy and Coriolis forces are negligible, and so although these forces
are crucial to driving and shaping the primary vortex, they play no direct
role in eye formation in a Boussinesq fluid. 
\end{abstract}

\section{Introduction}
\nocite{Pearce}
One of the most striking features of atmospheric vortices, such as tropical 
cyclones, is that they often develop a so-called eye; a region of reversed 
flow in and around the axis of the vortex. Much has been written about eye 
formation, particularly in the context of tropical cyclones, but the key 
dynamical processes are still poorly understood \citep{Pearce,Smith,Pearce2}. 
Naturally 
occurring vortices in the atmosphere are, of course, complicated objects, 
whose overall dynamics can be strongly influenced by, for example, planetary 
rotation, stratification, latent heat release through moist convection, and 
turbulent diffusion. Indeed the structure of eyes in tropical cyclones is 
almost certainly heavily influenced by both moist convection and
stratification. 
However, the ubiquitous appearance of eyes embedded within large-scale 
vortices suggests that the underlying mechanism by which they first form 
may be independent (partially if not wholly) of such complexities. 
Indeed eye-like structures are observed in other atmospheric vortices such as
tornadoes \citep[][and references therein]{Lugt} or polar lows
\citep{PolarLows}, which
are particularly interesting as they consist of large-scale
convective cyclonic structures observed in high latitudes polar regions.
To put the idea of a simple hydrodynamic mechanism
to the test we consider what is, perhaps, the simplest system in which 
eyes may form; that of steady axisymmetric convection in a rotating
Boussinesq fluid. We thus neglect the effects of stratification, and of 
moist convection.
Our underlying assumption is that some atmospheric phenomena could be
simple enough to be modelled in a uniform Boussinesq fluid. Indeed, a
recent study by \cite{Guervilly2014} noted that 
Boussinesq convection can yield, as in the atmosphere, the formation of 
large scale cyclonic vortices.

\begin{figure}
\centerline{
\includegraphics[width=0.7\textwidth]{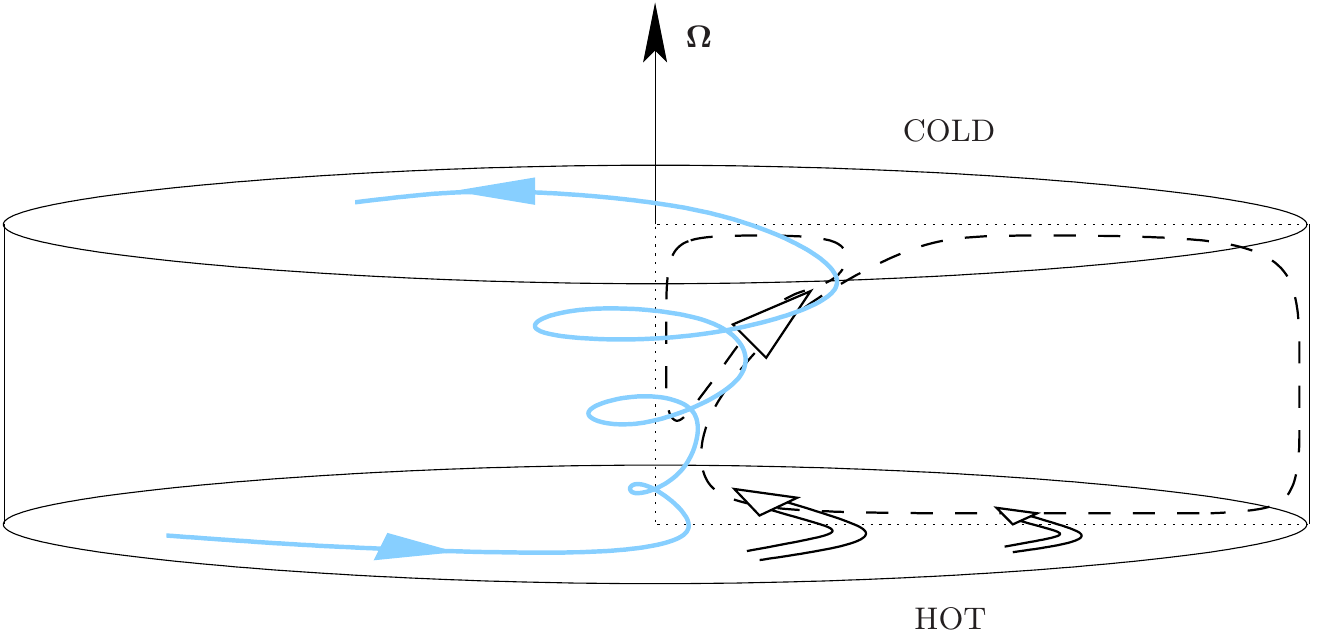}
}
\caption{Cartoon showing the global flow pattern in rotating
  convection. The motion in the vertical plane consists of the primary
  vortex, the eye-wall and the eye, while the azimuthal motion
  consists of regions of high angular momentum near the axis and low
  (or even negative) angular momentum at larger radii (the vertical axis is
  stretched by a coefficient $5$ for readability).}
\label{fig1}
\end{figure}

In this work, we 
consider a rotating, cylindrical domain in which
the lower surface is a no-slip boundary, the upper surface stress
free, and the motion driven by a prescribed vertical flux of heat. 
In a frame of reference rotating with the lower
boundary, the Coriolis force induces swirl in the convecting fluid,
which in turn sets up an Ekman-like boundary layer on the lower
surface. The primary flow in the vertical plane is then radially
inward near the lower boundary and outward at the upper surface. As
the fluid spirals inward, it carries its angular momentum with it
(subject to some viscous diffusion) and this results in a region of
particularly intense swirl near the axis. The overall flow pattern 
is as shown schematically in Figure~\ref{fig1}.

In the vertical plane the primary vortex has a
  clockwise motion, and so has positive azimuthal vorticity. If an eye
  forms, however, its motion is anticlockwise in the vertical plane
  (Figure~\ref{fig1}), and so the eye is associated with negative azimuthal
  vorticity. A key question, therefore, is: where does this negative
  vorticity come from? We shall show that it is not generated by
  buoyancy, since such forces are locally too weak. Nor does it arise
  from so-called vortex tilting, despite the local dominance of this
  process, because vortex tilting cannot produce any net azimuthal
  vorticity. Rather, the eye acquires its vorticity from the
  surrounding fluid by cross-stream diffusion, and this observation
  holds the key to eye formation in our simple system.

The region that separates the eye from the primary
  vortex is usually called the eyewall, and we shall see that this
  thin annular region is filled with intense negative azimuthal
  vorticity. So eye formation in our model problem is
  really all about the dynamics of creating an eyewall. In this paper
  we use numerical experiments to investigate the processes by which
  eyes and eyewalls form in our model system. We identify the key
  dynamical mechanisms and force balances, and provide a simple
  criterion which needs to be met for an eye to form.

\section{Problem Specification and Governing Equations}

\begin{figure}
\centerline{
\includegraphics[width=1\textwidth]{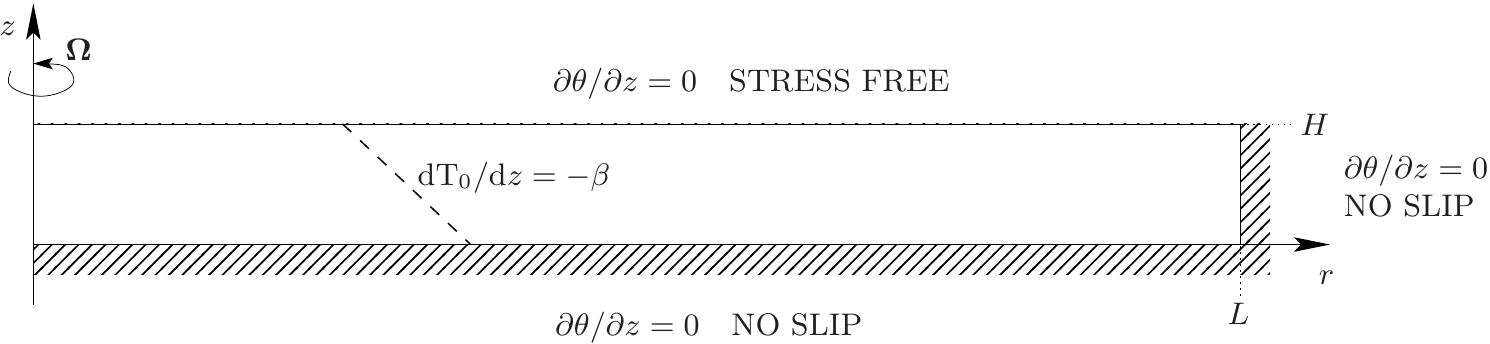}
}
\caption{Flow domain and boundary conditions.}
\label{fig2}
\end{figure}

We consider the steady flow of a Boussinesq fluid in a rotating, 
cylindrical domain of height $H$ and radius $R$, with $R \gg H$. The aspect
ratio is denoted as $\epsilon=H/R$. The flow is described in cylindrical
  polar coordinates, $(r,\phi,z)$, where the lower  
surface, $z=0$, and the outer radius, $r=R$, are no-slip boundaries. The 
upper surface, $z=H$, is impermeable but stress free. The motion is driven 
by buoyancy with a fixed upward heat flux maintained between the surfaces 
$z=0$ and $z=H$. In static equilibrium there is a uniform temperature
gradient, 
$\dd\T_0/\dd z=-\beta$. We decompose $\T=\T_0(z)+\theta$, 
where $\theta$ is the perturbation in temperature from the linear profile. 
In order to maintain a constant heat flux the thermal boundary conditions 
on the surfaces  $z=0$ and $z=H$ are $\partial \theta/\partial z=0$, while 
the outer radial boundary is thermally insulating, $\partial
\theta/\partial r=0$. 
The flow domain and boundary conditions are summarised in Figure~\ref{fig2}.

We adopt a frame of reference that rotates with the boundaries.
Denoting $\bfOmega$ the background rotation rate, $\bfg$ the gravitational acceleration,
$\nu$ the kinematic viscosity of the fluid, 
$\kappa$ its thermal diffusivity, and $\alpha$ its thermal expansion
coefficient, the governing equations are 
\begin{multiequations}
\label{eq_NS}
\begin{equation}
\frac{\bigD \bfu}{\bigD t}
=
- \frac{1}{\rho_0}\, \bfnabla p 
- 2 \bfOmega \times \bfu
+ \nu \, \bfnabla ^2 \bfu
-
\alpha 
\, \theta \, \bfg  \, , \qquad \bfnabla \cdot \bfu=0 \, ,
\end{equation} 
\singleequation
and
\begin{equation}
\frac{\bigD \theta}{\bigD t}
=
\kappa
 \, \bfnabla ^2 \theta
+ \beta \, \uu_z \,,
\label{eq_theta}
\end{equation}
\end{multiequations} 
\citep[e.g.][]{Chandra,Drazin}. 
We further restrict ourselves to axisymmetric motion, so that we may
decompose the velocity field into poloidal and azimuthal velocity
components, $\bfu_p=(\uu_r,0,\uu_z)$ and $\bfu_\phi=(0,\uu_\phi,0)$,
which are separately solenoidal. The azimuthal component of (\ref{eq_NS}$a$)
then becomes an evolution equation for the specific angular momentum in the
rotating frame, $\Gamma=r\uu_\phi$, 
\begin{equation}
\frac{\bigD \Gamma}{\bigD t} =  - 2 r \,\Omega\, \uu_r + \nu \stokes(\Gamma)\,,
\label{Gammadim}
\end{equation} 
where
\[
\stokes=r \frac{\partial }{\partial r}\left(\frac{1}{r}\frac{\partial }{\partial r}\right) + \frac{\partial^2 }{\partial z^2}
\]
is the Stokes operator.
Moreover the curl of the poloidal components yields an evolution equation for
the azimuthal vorticity, $\bfomega_\phi= \bfnabla \times \bfu_p \, ,$
\begin{equation}
\frac{\bigD}{\bigD t}
\left(\frac{\omega_\phi}{r}\right)=\frac{\partial}{\partial
  z}\left(\frac{\Gamma^2}{r^4}\right) + \frac{2\, \Omega}{r} \frac{\partial
  \uu_\phi}{\partial z}  -\frac{\alpha g}{r}\,   \frac{\partial
  \theta}{\partial r} + \frac{\nu}{r^2}\stokes\left(r \omega_\phi \right)
\,. 
\label{omegadim}
\end{equation}
We recognize the curl of the Coriolis, buoyancy and viscous forces on the 
right of (\ref{omegadim}). The axial gradient in $\Gamma^2/r^4$ is, perhaps, a 
little less familiar as a source of azimuthal vorticity. However, this arises 
from the contribution of $\bfnabla \times \left( \bfu_\phi \times \bfomega_p 
\right)\, , $ where $\bfomega_p =\bfomega - \bfomega_\phi \, ,$  to the
vorticity equation and represents the self-advection  
(spiralling up) of the poloidal vorticity-lines by axial gradients in swirl
\citep[e.g.][]{Davidson}.
The scalar equations (\ref{Gammadim}) 
and (\ref{omegadim}) are formally equivalent to (\ref{eq_NS}$a$), with $\Gamma$ 
and $\omega_\phi$ uniquely determining the instantaneous velocity 
distribution. Finally, it is convenient to introduce the Stokes 
stream-function, $\psi$, which is defined by $\bfu_p= \bfnabla \times 
\left[\left(\psi/r\right) \bfe_\phi\right]$ and related to the azimuthal 
vorticity by
$
r\omega_\phi=-\stokes \psi\,.
$

\section{Global Dynamics}
We perform numerical simulations in the form of an initial value
problem which is run until a steady state is reached. We solve
equations (\ref{eq_theta}), (\ref{Gammadim}) and (\ref{omegadim})
in which the length has been scaled with the height $H$ of the system,
the time with $\Omega ^{-1}\, ,$ 
and the temperature with $H \beta\,.$ 
The dimensionless control parameters of the system are the Ekman number
\(
\Ek={\nu}/{\Omega H^2}\,,
\)
the Prandtl number
\(
\Pra={\nu}/{\kappa}\,,
\)
and the Rayleigh number
\(
\Ray={\alpha g \beta H^4}/{\nu \kappa}\,.
\)

We use second-order finite differences and an implicit second-order
backward differentiation (BDF2) in time. The number of radial and
axial cells is $1000\times500$, and in each simulation grid resolution
studies were undertaken to ensure numerical convergence.
The aspect ratio of the computational
domain is set at $\epsilon=0.1$, a ratio inspired by tropical storms
(for which $H \simeq 10$km, and $R \simeq 100$km).  The Ekman number $\Ek$
is set to $0.1$, which is a sensible turbulent estimate for tropical
cyclones. The values of $\Pr$ and $\Ray$ will be varied through this study
to control the strength of the convection.

We shall consider flows in which the local Rossby number 
$\Ro=\uu_\phi/{\Omega H}\,,$ is of the order
unity or 
less at large radius, $r \simeq L \, ,$ 
but is large near the axis, $r \simeq H \, ,$ which is not
untypical of a tropical cyclone and turns out to be the regime in which an eye and
eyewall form in our numerical simulations. We shall also take a suitably defined
Reynolds number, $\Ree \,,$ to be considerably larger than unity, though not so large that
the laminar flow becomes unsteady. A moderately large Reynolds number also turns
out to be crucial to eye formation.

\begin{figure}
\centerline{
\includegraphics[width=1\textwidth]{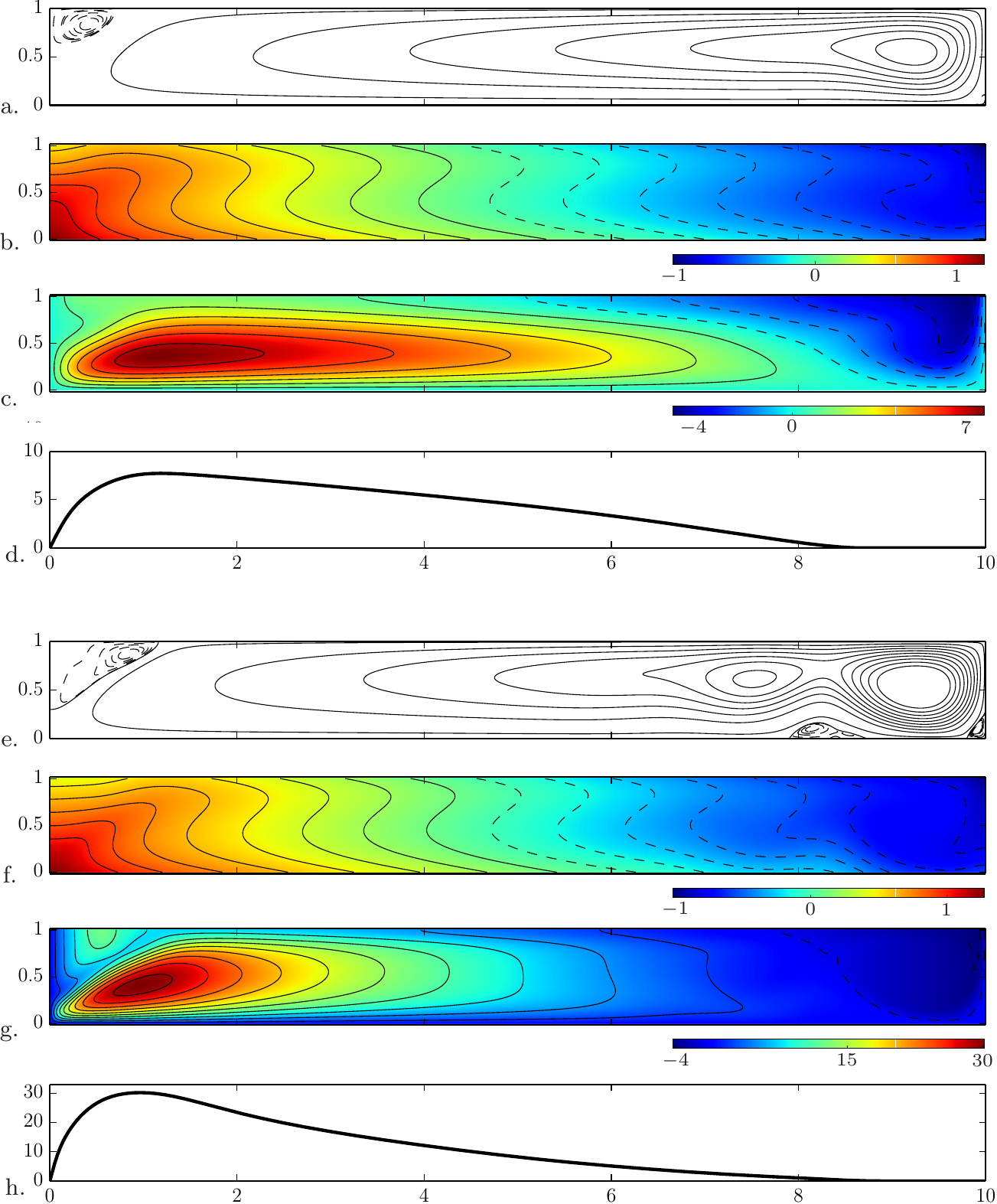}
}
\caption{Steady state solution in the $(r,z)$-plane for the parameters $\Pra=0.5$ and $\Ray=1.5\times 10^{4}$ in (a--d), and for $\Pra=0.1$ and $\Ray=2 \times 10^{4}$ in (e--h).
(a,e) The stream-function distribution, 
(b,f) the total temperature $\T=\T_0(z)+\theta$,
(c,g) the azimuthal velocity, $\uu_\phi$, 
(d,h) the radial variation of $\Ro$.} 
\label{fig3}
\end{figure}

In order
to focus thoughts, let us start by considering two specific cases:
$\Pr = 0.5\, , \,\,\Ray = 1.5\times 10^{4}\, ,$ and
$\Pr = 0.1\, , \,\,\Ray = 2\times 10^{4}\, .$ These two cases are
represented in Fig.~\ref{fig3}.
In both cases, the primary flow in the vertical plane is radially inward near the lower
boundary and outward at the upper surface.
The steady state stream-function distributions are shown in
Fig.~\ref{fig3}a,e.
It is evident that in both
cases an eye
has formed near the axis, but it is much more pronounced in the later case. 
Fig.~\ref{fig3}b,f show the corresponding
distributions of total temperature. The poloidal flow sweeps heat towards
axis at low values of $z$, 
causing a build-up of heat near the axis with a corresponding cooler
region at larger radii. The resulting negative radial gradient in $\T$
drives the main poloidal vortex, ensuring that it has positive
azimuthal vorticity in accordance with (\ref{omegadim}).
Fig.~\ref{fig3}c,g present the distributions of the azimuthal velocity.
The Coriolis force induces swirl in the convecting fluid. As the 
fluid spirals inward, it carries its angular momentum with it (subject to 
some viscous diffusion) and this results in a region of particularly intense 
swirl near the axis.
It also shows a substantial region of negative (anti-cyclonic rotation)
at large radius, something that is also observed in tropical cyclones
\citep[e.g.][]{Frank77}.  In order to quantify the strength of the azimuthal
flow, we introduce a Rossby number which is a function of radius
$\Ro(r)={\left(\uu_\phi\right)_{\rm max}}/{\Omega
  H}\,,$ where
$\left(\uu_\phi\right)_{\rm max}$ is the maximum value of azimuthal
velocity at any one radial location. The Rossby number as a function of
radius is represented in Fig.~\ref{fig3}d,h.  The eye
is obtained when the local Rossby number, is of the order unity at
large radius, $r\simeq L$, but is large near the axis,
$r\simeq H$.
Given the increase of $\Ro$ near the axis, the 
Coriolis force can be neglected in the vicinity of the eye. 
Note that, in the second
case (Fig.~\ref{fig3}e,f,g,h), the Rossby number is much larger, and the
eye much more pronounced than in the first case (Fig.~\ref{fig3}a,b,c,d).

To summarise, in both cases, the buoyancy force evidently drives motion in the poloidal plane,
which in turn induces spatial variations in angular momentum,
$\Gamma$, through the Coriolis force, $2 \Omega r\uu_r$, in
(\ref{Gammadim}). 
The flow spirals radially inward along the lower
boundary and outward near $z=H$, as shown in Fig.~\ref{fig1}. The Coriolis force then ensures that 
the angular momentum, $\Gamma$, 
rises as the fluid spirals inward along the bottom boundary, but falls
as it spirals back out along the upper surface towards
$r=L$. 
The swirl of the flow as it approaches the axis is thus controled both by
the Ekman number and by the aspect ratio of the domain.
With our choice of the parameters, particularly high levels of azimuthal
velocity built up near the axis, with a correspondingly large value of
$\Ro$ in the vicinity of the eyewall.  In some sense, then, the global
flow pattern is both established and shaped by the buoyancy and
Coriolis forces, yet, as we shall discuss, these forces are negligible in
the viscinity of the eye.

\section{Global Versus Local Dynamics}
\subsection{The anatomy of eyewall formation}
\label{ana}

The large value of $\Ro$ near the axis means that the Coriolis force
is locally negligible in the region where the eye and eyewall form,
and it turns out that this is true also of the buoyancy force in our
Boussinesq simulations. Thus the very forces that establish the global
flow pattern play no significant role in the local dynamics of the
eye. It is worth considering, therefore, the simplified version of
(\ref{Gammadim}) and (\ref{omegadim}) which operate near the axis,
\begin{equation}
\frac{\bigD \Gamma}{\bigD t} \simeq  \nu \stokes(\Gamma)\,, 
\label{sys1a}
\end{equation}
and
\begin{equation}  
\frac{\bigD}{\bigD t} \left(\frac{\omega_\phi}{r}\right) \simeq
\frac{\partial}{\partial z}\left(\frac{\Gamma^2}{r^4}\right) 
+ \frac{\nu}{r^2}\stokes\left(r \omega_\phi \right)\,.
\label{sys1b}
\end{equation}

\begin{figure}
\centerline{
\includegraphics[width=0.8\textwidth]{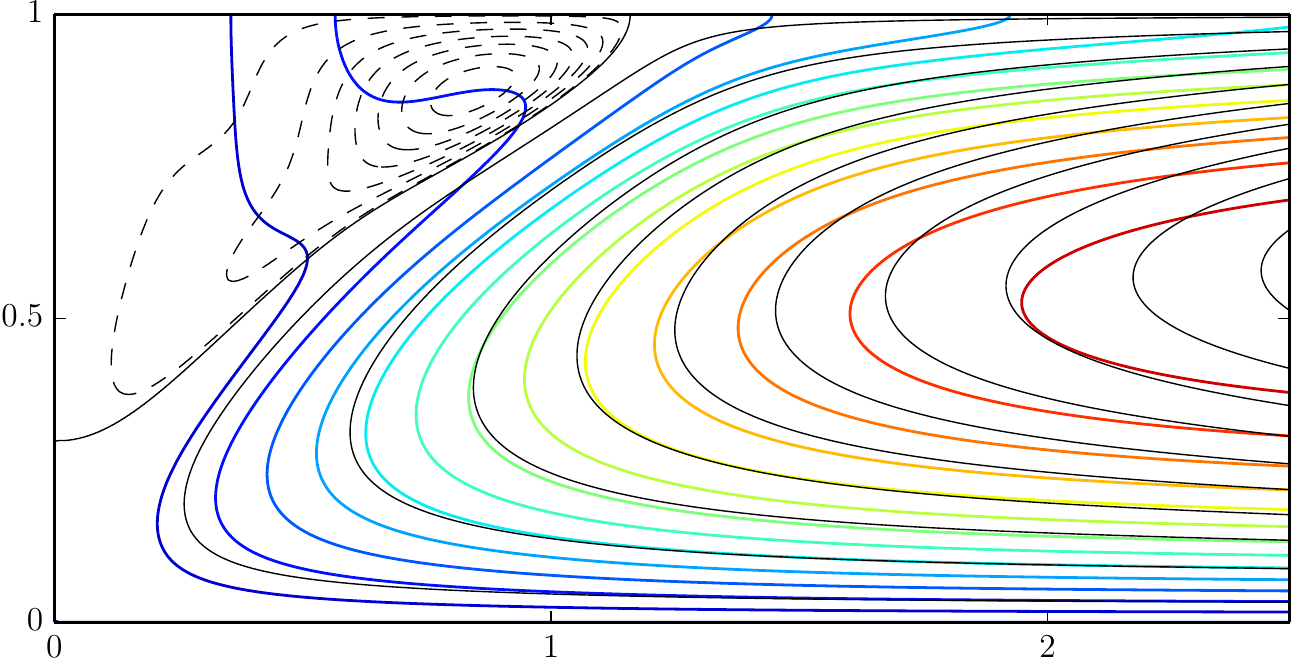}
}
\caption{Contours of constant angular momentum (color) superimposed on the
streamlines (black) in the inner quarter of the flow domain
($(r,z)$-plane), for $\Pr=0.1$ and $\Ray=2\times 10^4$.}
\label{fig4}
\end{figure}

The eye is characterised by anticlockwise motion in the $(r,z)$-plane 
($\omega_\phi<0$), in contrast to the global vortex that is clockwise 
($\omega_\phi>0$). It is also characterised by low levels 
of angular momentum. 
A natural question to ask, therefore, is where this 
negative azimuthal vorticity comes from. Since $\Gamma$ is small in the eye, 
$\omega_\phi/r$ is locally governed by a simple advection diffusion equation 
in which the source term is negligible, and so the negative azimuthal 
vorticity in the eye has most probably diffused into the eye from the eyewall. 
This kind of slow cross-stream diffusion of vorticity into a region of closed 
streamlines is familiar from the Prandtl-Batchelor theorem \citep{Batchelor}, and in 
this sense the eye is a passive response to the accumulation of negative 
$\omega_\phi$ in the eyewall. If this is substantially true, and we shall 
see that it is, then the key to eye formation is the generation of significant 
levels of negative azimuthal vorticity in the eyewall, and so the central 
questions we seek to answer is how, and under what conditions, the eyewall 
acquires this negative vorticity.

Given that the Reynolds number is large, it is tempting to consider the 
inviscid limit and attribute the growth of negative $\omega_\phi$ to the 
first term on the right of (\ref{sys1b}). That is, axial gradients in $\Gamma$ 
can act as a local source of azimuthal vorticity, and indeed this mechanism 
has been invoked by previous authors in the context of tropical cyclones 
\citep[e.g.][]{Pearce,Smith,Pearce2}.
The idea is that, in steady state, 
if viscous diffusion is ignored in the vicinity of the eyewall, (\ref{sys1a}) 
and (\ref{sys1b}) locally reduce to 
\begin{equation}
\Gamma=\Gamma\left(\psi\right)\, ,
\label{sq1a}
\end{equation} 
and
\begin{equation} 
\bfu \cdot \bfnabla\left(\omega_\phi/r\right)=\frac{\partial}{\partial
  z}\left(\frac{\Gamma^2}{r^4}\right)=-2\frac{\Gamma
  \Gamma^\prime\left(\psi\right)}{r^3}\uu_r \, .
\label{sq1b}
\end{equation}

To the extent the viscosity can be ignored,
$\Gamma\left(\psi\right)$ increases to a maximum at roughly mid-height,
where $\psi$ is a maximum,  
and then drops off as we approach the upper boundary. Thus
$\Gamma^\prime\left(\psi\right)>0$, and so according to (\ref{sq1b})
positive vorticity is induced  
as the streamlines curve inward and upward, while negative vorticity is 
created after the streamlines turn around and $\uu_r$ reverses sign. 
Since the eyewall is associated with the upper region, where the flow is outward, 
it is natural to suppose that the negative vorticity in the eyewall arises 
from precisely this process. 
However, in the case of a Boussinesq fluid, this term cannot produce 
any net negative azimuthal vorticity in the eyewall, essentially because 
the term on the right of (\ref{sq1b}) is a divergence. To see why this is so, 
we rewrite (\ref{sq1b}) as 
\begin{equation} 
\bfnabla \cdot \left[\left(\omega_\phi/r\right)\bfu\right]=\bfnabla \cdot
\left[\left(\Gamma^2/r^4\right)\bfe_z\right]\,, 
\end{equation} 
and integrate this over a control volume in the form of a stream-tube in 
the $r-z$ plane composed of two adjacent streamlines that pass through 
the eyewall. If the stream-tube within the control volume starts and ends 
at a fixed radius somewhat removed from the eyewall, then the right-hand 
divergence integrates to zero. The flux of vorticity into the control 
volume (the stream-tube) is therefore the same as that leaving.
 In short, the azimuthal 
vorticity generated in the lower regions where the streamlines curve inward 
and upward is exactly counterbalanced by the generation of negative vorticity 
in the upper regions where the flow is radially outward. Such a process 
cannot result in any negative azimuthal vorticity in the eyewall.
Returning to (\ref{sys1b}) we conclude that the only possible source of net 
negative vorticity in the eyewall is the viscous term, and this drives us 
to the hypothesis that the negative vorticity in the eyewall has its origins 
in the lower boundary layer. That is, negative azimuthal vorticity is 
generated at the lower boundary and then advected up into eyewall where it 
subsequently acts as the source for a slow cross-stream diffusion of negative 
vorticity into the eye. Of course, as the streamlines pass up into, and then 
through, the eyewall there is also a generation of first positive and then 
negative azimuthal vorticity by the axial gradients in $\Gamma^2/r^4$, but 
these two contributions exactly cancel, and so cannot contribute to the 
negative vorticity in the eyewall.

\begin{figure}
\centerline{
\includegraphics[width=1.0\textwidth]{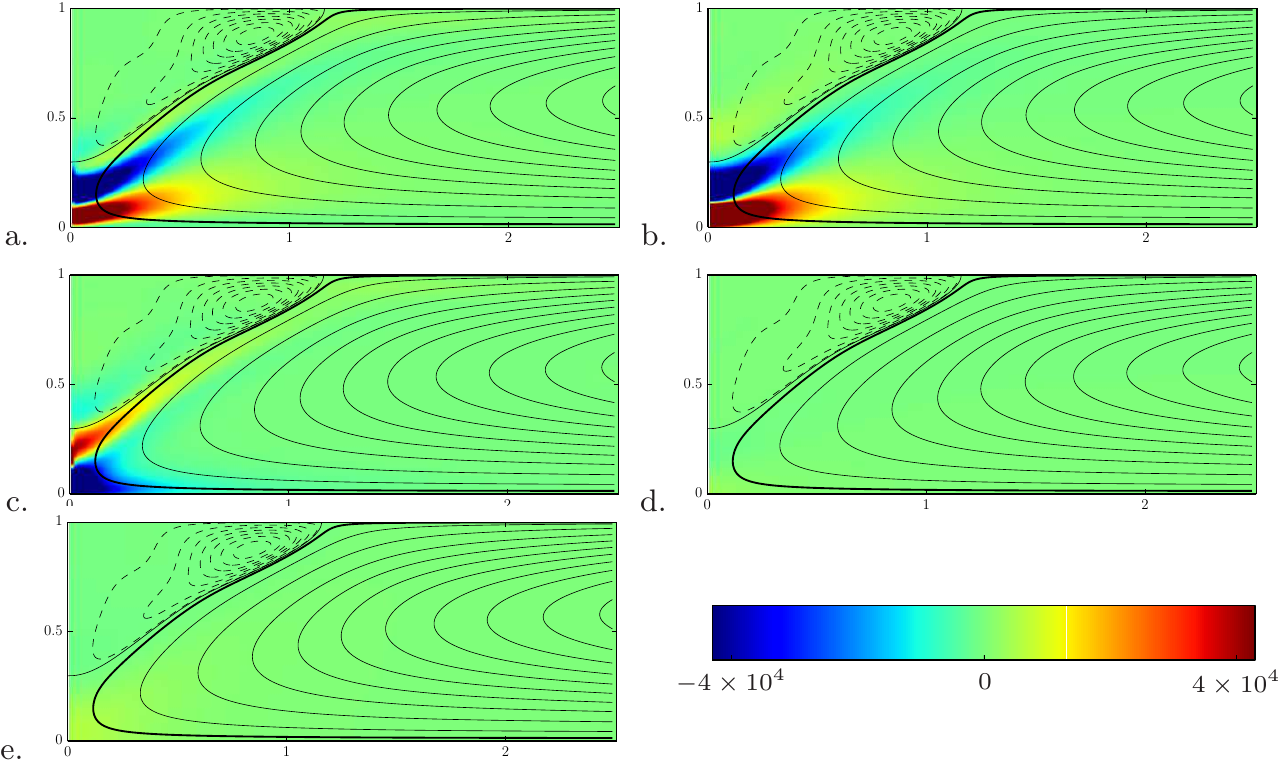} 
}
\caption{Colour maps of the distribution of the
    various forces in the azimuthal vorticity equation (\ref{omegadim}) in the inner
    part of the flow domain ($(r,z)$-plane): (a) the convective derivative of
    azimuthal vorticity; (b) the term associated with axial gradients in
    $\Gamma$; (c) diffusion; (d) the Coriolis
    term; and (e) buoyancy. Parameters: $\Pr=0.1\, ,\ \Ray=2\times 10^4$.}
\label{fig5}
\end{figure}

\begin{figure}
\centerline{
\includegraphics[width=0.8\textwidth]{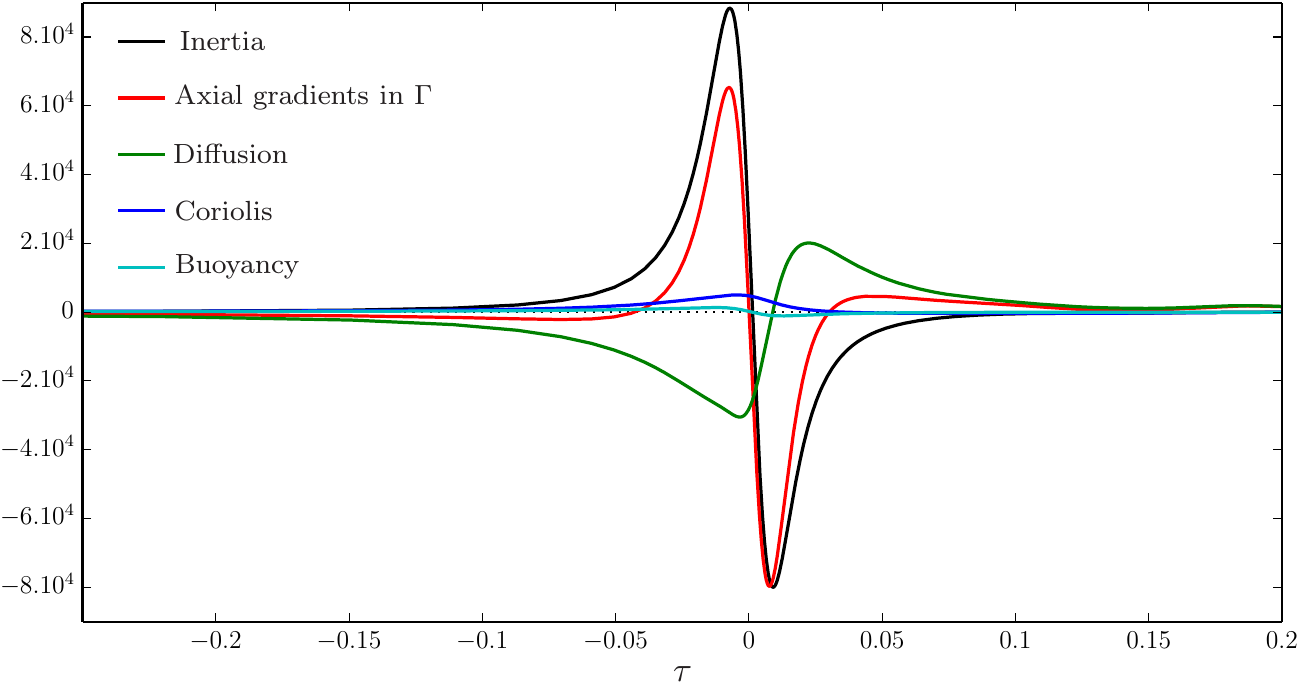}}
\caption{The variation of the terms in the azimuthal vorticity equation 
(\ref{omegadim}) with position on the streamline that passes through 
the centre of the eyewall. The convective derivative on the left-hand side 
of (\ref{omegadim}) is black, the term associated with axial gradients in 
$\Gamma$ is red, the viscous term is green, and the Coriolis and buoyancy 
terms (blue and light blue) are indistinguishable from the $x$-axis.
 Parameters: $\Pr=0.1\, ,\ \Ray=2\times 10^4$.}
\label{fig6}
\end{figure}

\begin{figure}
\centerline{
\includegraphics[width=1\textwidth]{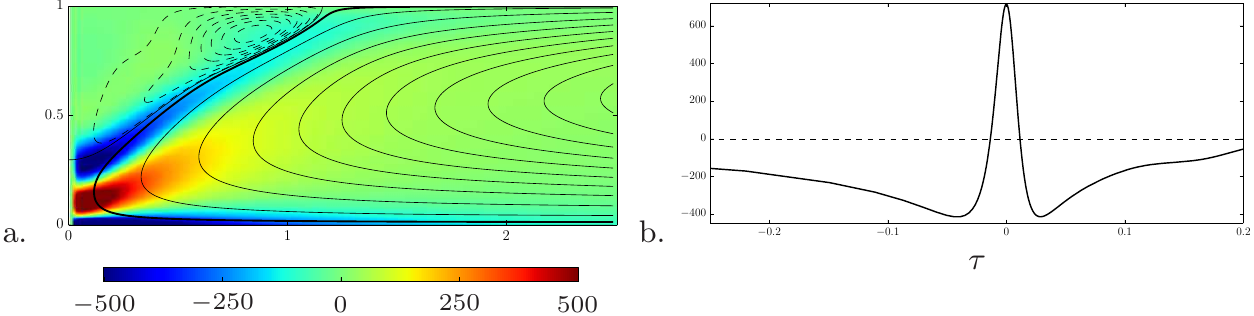}
} 
\caption{(a) Colour map of $\omega_\phi/r$ superimposed on the streamlines
  in the inner quarter of the domain ($(r,z)$-plane). (b) The variation of $\omega_\phi/r$ along 
the streamline that passes through the centre of the eyewall (as indicated 
by the thick black line in a). Parameters: $\Pr=0.1\, , \ \Ray=2 \times 10^4$.}
\label{fig7}
\end{figure}

The above description is put to the test in Fig.~\ref{fig4} to Fig.~\ref{fig7}. 
A more detailed view of the angular momentum 
distribution and streamlines in the region adjacent to the eye and eyewall
is provided in Fig.~\ref{fig4}.
In the region to the right of the eye, the contours of 
constant angular momentum are roughly aligned with the streamlines, 
in accordance with (\ref{sq1a}). Note that, although the contours of 
constant $\Gamma$ roughly follow the streamlines, the two sets of contours 
are not entirely aligned. This is mostly a result of cross-stream 
diffusion, particularly near the eyewall, although it also partially 
arises from the (relatively weak) Coriolis force. 

Fig.~\ref{fig5} shows the distribution and relative magnitudes of the various forces in the
azimuthal vorticity equation across the inner part of the flow
domain. 
The buoyancy and Coriolis terms, though important
for the large scale dynamics, are locally negligible (panels
\ref{fig5}(d) and \ref{fig5}(e)), while diffusion is largely limited
to the boundary layer, the eyewall, and a region near the axis where
the flow turns around (panel \ref{fig5}c). Note also that
$\partial\left(\Gamma^2/r^4\right)/\partial z$ is small within the eye
(panel \ref{fig5}b) but there are intense regions of equal and
opposite $\partial\left(\Gamma^2/r^4\right)/\partial z$ below the
eyewall, which are matched by corresponding regions of equal and
opposite $\bfu \cdot \bfnabla\left(\omega_\phi/r\right)$ in panel
\ref{fig5}a. These figures appear to validate that (\ref{sys1b}) is
the relevant equation in this domain. It is interesting that the very
forces that shape the global flow, i.e.~the Coriolis and buoyancy forces,
play no role in the vicinity of the eye. 

This can be further highlighted by introducing
the variable $\tau$, defined as a parametric coordinate along 
an iso-$\psi$. It corresponds to the position at a given time $\tau$ for 
a particle advected along a streamline; it stems from
\(
\left(\left.{\dd z}/{\dd \tau}\right)\right|_{\psi={\rm cst}}=\uu_z\,.
\)
The constant of integration is set such that $\tau= 0$ for the maximum 
of $\omega_\phi/r$. 
Fig.~\ref{fig6} shows 
forces in the azimuthal vorticity equation, 
as a function of position on the streamline that passes 
through the centre of the eyewall (thick streamline in Fig.~\ref{fig5}). 
To first order there is an approximate 
balance between the advection of $\omega_\phi/r$  and
$\partial\left(\Gamma^2/r^4\right)/\partial z$, though there is a
significant contribution from  
the diffusion of vorticity within the eyewall. Both the Coriolis and 
buoyancy terms are completely negligible. In short, the force balance 
is that of (\ref{sys1b}). 

The main features of the eyewall are most clearly seen in Fig.~\ref{fig7}a, 
which shows the distribution of azimuthal vorticity, $\omega_\phi/r$, 
superimposed  on the streamlines. The exceptionally strong levels of 
azimuthal vorticity in and around the eyewall is immediately apparent, and
indeed it is tempting to define the eyewall as the outward sloping region
of strong negative azimuthal vorticity which separates the eye from the
primary vortex.  
There are two other important features of Fig.~\ref{fig7}a. First, a large 
reservoir of negative azimuthal vorticity builds up in the lower boundary 
layer, as it must. Second, between the lower boundary and the eyewall there 
is a region of intense positive azimuthal vorticity. 
Fig.~\ref{fig7}b. shows 
the variation of $\omega_\phi/r$ along the streamline that passes through 
the centre of the eyewall, as indicated by the thick black line in Fig.~\ref{fig7}a. 
As the streamline passes along the bottom boundary 
layer, $\omega_\phi/r$ becomes progressively more negative. There is then 
a sharp rise in $\omega_\phi/r$  as the streamline pulls out of the boundary 
layer and into a region of positive $\partial \Gamma/\partial z$, followed by 
a corresponding drop as the streamline passes into the region of negative
$\partial \Gamma/\partial z$. Crucially, the rise and subsequent fall in
$\omega_\phi/r$ caused by axial gradients in angular momentum exactly
cancel, and  
so the fluid emerges into the eyewall with the same level of vorticity it 
had on leaving the boundary layer.

\begin{figure}
\centerline{
\includegraphics[width=0.8\textwidth]{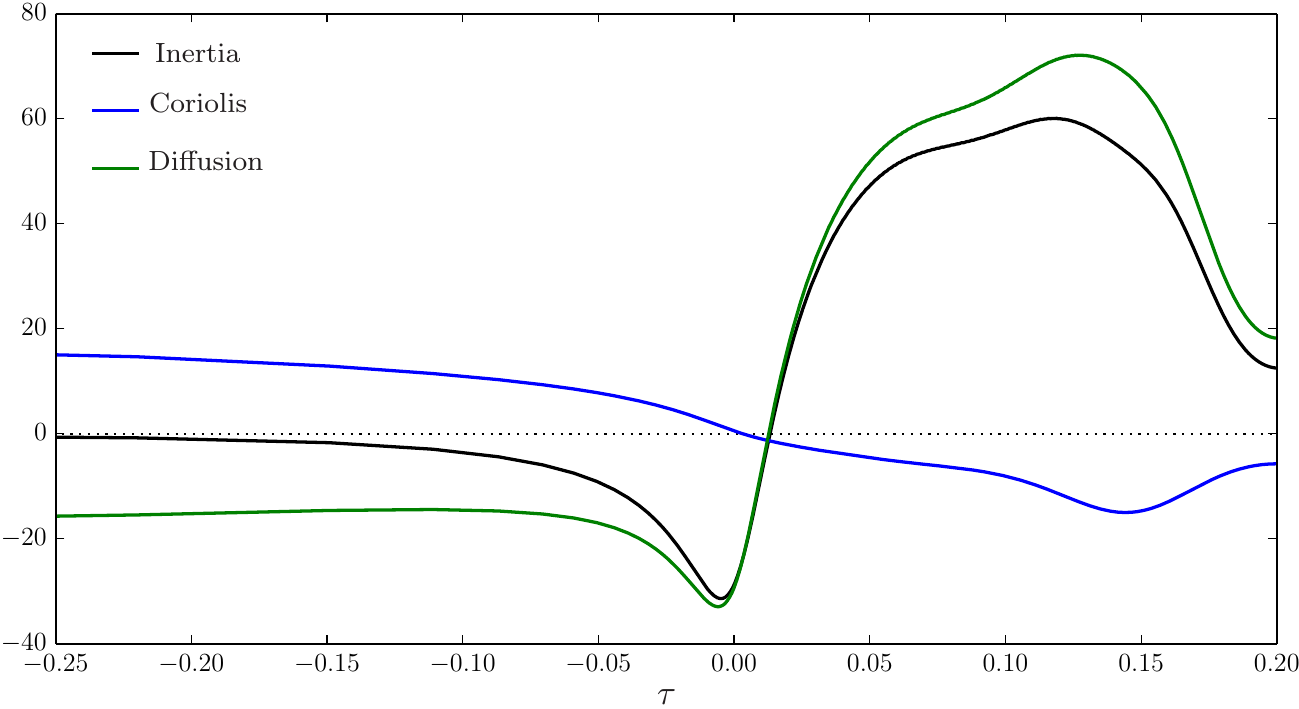}}
\caption{The variation of the various torques in the angular momentum
  equation \eqref{Gammadim} with position on the streamline that passes through the
  centre of the eyewall. The convective derivative on the left-hand side of
  \eqref{Gammadim} is black, the viscous diffusion term is green, and the Coriolis term
  is blue.}
\label{fig8}
\end{figure}

\begin{figure}
\centerline{
\includegraphics[width=1\textwidth]{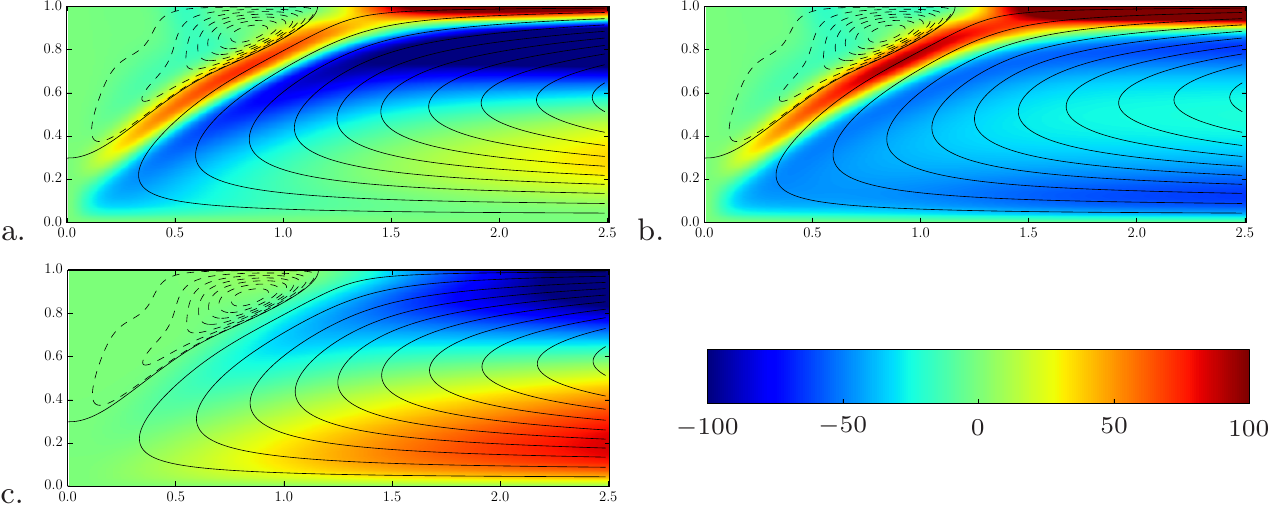}
}
\caption{The spatial distribution of the various torques
  in the angular momentum equation in the inner
    part of the flow domain ($(r,z)$-plane) : (a) the convective
  derivative of $\Gamma$, (b) the diffusion term on the right,
  and (c) the Coriolis torque. Parameters: $\Pr=0.1\, , \  \Ray=2\times 10^4$.}
\label{fig9}
\end{figure}

Finally we consider in Fig.~\ref{fig8} and Fig.~\ref{fig9} the distribution 
of the various contributions to the angular momentum equation 
(\ref{Gammadim}).
Fig.~\ref{fig8} shows the variation of these terms along the streamline that
passes through the centre of the eyewall, as indicated by the thick black
line in Figure~\ref{fig5}. As before, the origin for the horizontal axis is taken
to be the innermost point on the streamline. Clearly, within the eyewall
there is a leading-order balance between the advection and diffusion of
angular momentum, in accordance with the approximate equation~\eqref{sys1a}, although
the Coriolis torque is not entirely negligible. However, within the
boundary layer ahead of the eyewall the force balance is quite different,
with the Coriolis and viscous forces being in approximate balance, the
convective growth of angular momentum being small.   

A similar impression may be gained from Fig.~\ref{fig9} which shows, for the
inner quarter of the flow domain, the distribution of the various
contributions to equation~\eqref{Gammadim}. 
The panel (a) 
is the convective derivative of $\Gamma$ on the left of (\ref{Gammadim}), the panel (b) 
is the diffusion term on the right, and the panel (c) is the Coriolis torque. 
Clearly, the positive Coriolis torque acting near the bottom boundary
layer is largely matched by local cross-stream diffusion, the convective 
growth of angular momentum being small. The
negative Coriolis torque acting on the outflow near the upper surface
is mostly balanced by $\bfu \cdot \bfnabla \Gamma$, resulting in a fall in $\Gamma$.
Within the eyewall, the force balance is quite different, since there is 
a leading-order balance between the advection and
diffusion of angular momentum, and the Coriolis torque is very weak. 
This confirms our previous observations in Fig.~\ref{fig5}, and is in 
accordance with the approximate equation 
(\ref{sys1a}).

To summarise, in this particular simulation the eyewall that separates
the eye from the primary vortex is characterised by high levels of
negative azimuthal vorticity. That vorticity comes not from the term 
$\partial(\Gamma^2/r^4)/\partial z$,
despite its local dominance, but rather from the boundary layer at $z =
0$. The eye then acquires its negative vorticity by cross-stream
diffusion from the eyewall, in accordance with the Prandtl-Batchelor
theorem. Although the global flow is driven and shaped by the buoyancy
and Coriolis forces, these play no significant dynamical role in the
vicinity of the eye and eyewall. As we shall see, these dynamical
features characterise all of our simulations that produce eyes.

\subsection{A comparison of vortices that do and do not form eyes}

\begin{figure}
\centerline{
\includegraphics[width=1\textwidth]{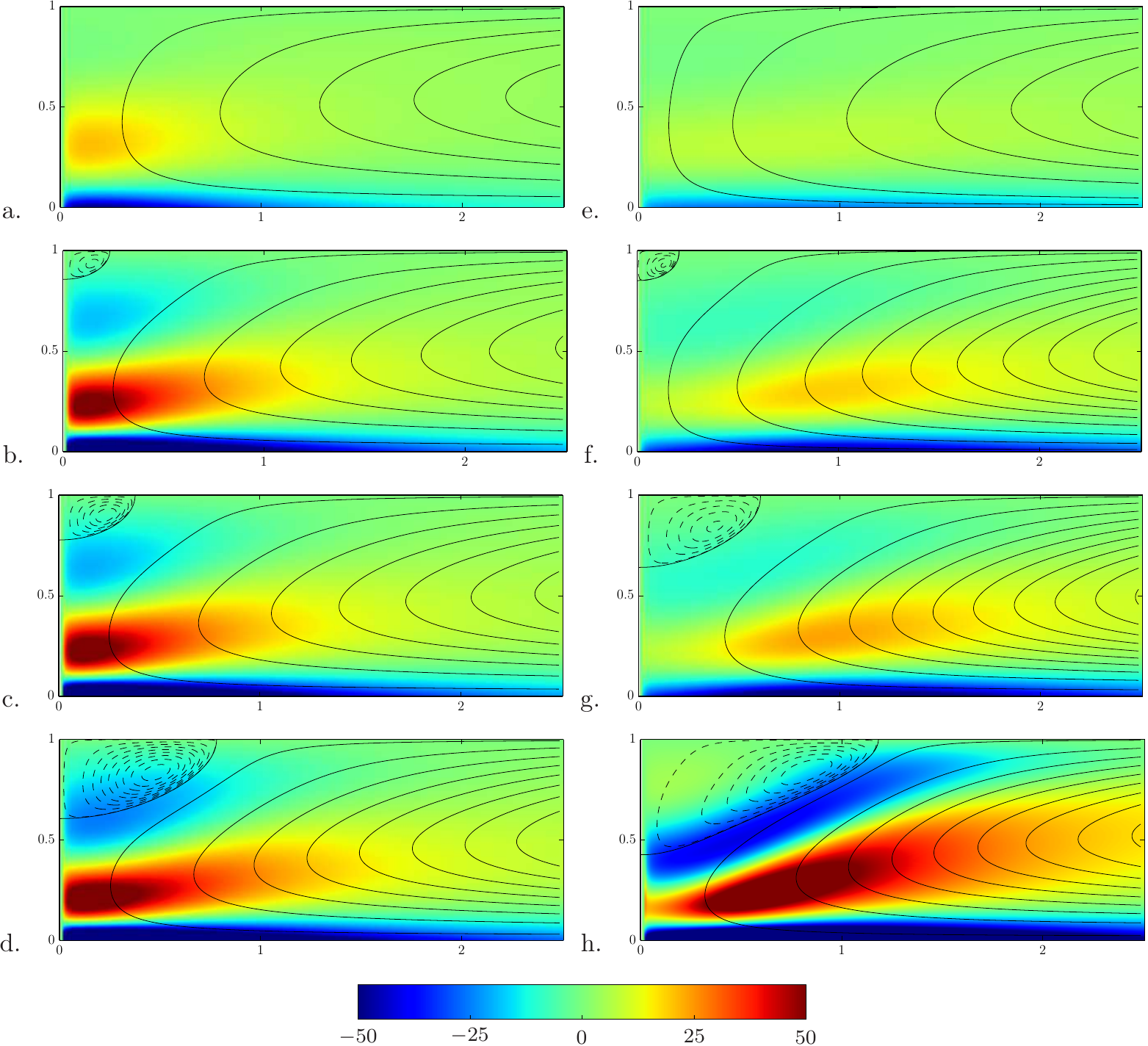}
}
\caption{Colour map of $\omega_\phi/r$ superimposed on the streamlines in
  the inner quarter of the domain ($(r,z)$-plane), for $\Pra=0.3$ and (a)
  $\Ray=2\times 10^{3}$, (b) $\Ray=5.5\times 10^{3}$, (c) $\Ray=6\times 10^{3}$, (d)
  $\Ray=9 \times 10^{3}$, and for  $\Pra=0.1$ and (e)
  $\Ray=10^{3}$, (f) $\Ray=1.7\times 10^{3}$, (g) $\Ray=2\times 10^{3}$, (h)
  $\Ray=5 \times 10^{3}\,.$ The colour code is the same on all graphs.} 
\label{fig10}
\end{figure}

\begin{figure}
\centerline{
\includegraphics[width=0.7\textwidth]{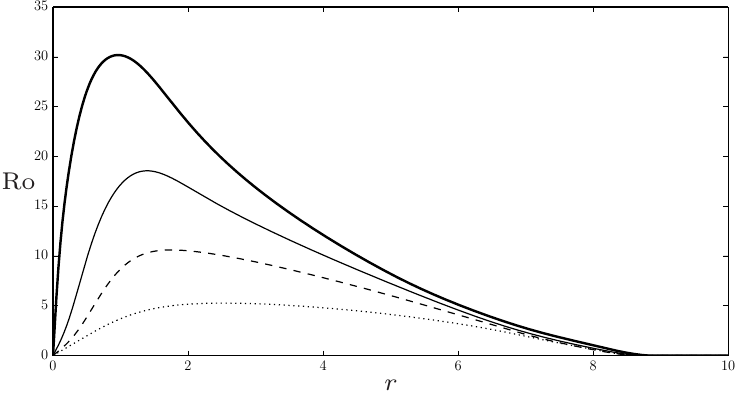}
}
\caption{The radial variation of $\Ro$, for the cases with $\Pr=0.1$ and 
$\Ray=10^{3}$ (dotted),  $\Ray=2\times 10^{3}$ (dashed), $\Ray=5\times 10^{3}$ 
(thin solid) and $\Ray=2 \times 10^{4}$ (thick solid).} 
\label{fig11}
\end{figure}

Let us now compare numerical simulations at given 
values of the Prandtl number ($\Pr = 0.3$ and $\Pr = 0.1$) and varying Rayleigh numbers. 
Figure~\ref{fig10} shows the stream-function distribution.
Clearly, eyes form in Fig.~\ref{fig10}b,c,d,f,g,h. 
From Table~1 and Fig.~\ref{fig11}, we see that the peak value of $\Ro$ is
of the order of $10$ or larger
in those cases where an eye forms, which is typical of an atmospheric 
vortex.

\begin{table}
\caption{Sample of numerical results, $\Ree$ is defined by equation \eqref{Re}.}
\centering
\begin{tabular}{rrrr}
\hline
$\Ray $\ \   &\hskip 1cm max$(\Ro)\!\!\! $ & \hskip 1cm $\Ree $\ \ \ & \hskip 1cm Eye?  \\
\hline
\multicolumn{4}{c}{$\epsilon = 0.1$ \ $\Ek = 0.1$ \ $\Pr=0.1$}\\
1000	&	5.27	&	17.78	&	0\\
1500	&	8.32	&	29.93	&	0\\
1650	&	9.07	&	33.2	&	0\\
1700	&	9.3	&	34.27	&	1\\
1750	&	9.53	&	35.32	&	1\\
1800	&	9.76	&	36.37	&	1\\
2000	&	10.61	&	40.44	&	1\\
5000	&	18.56	&	86.02	&	1\\
20000	&	30.18	&	178	&	1\\[1mm]
\multicolumn{4}{c}{$\epsilon = 0.1$ \ $\Ek = 0.1$ \ $\Pr=0.3$}\\
2000	&	3.39	&	18.98	&	0\\
4000	&	5.79	&	30.48	&	0\\
5000	&	6.72	&	34.75	&	0\\
5500	&	7.14	&	36.66	&	1\\
6000	&	7.54	&	38.44	&	1\\
8000	&	8.96	&	44.59	&	1\\
9000	&	9.59	&	47.17	&	1\\[1mm]
\multicolumn{4}{c}{$\epsilon = 0.1$ \ $\Ek = 0.1$ \ $\Pr=0.5$}\\
6000	&	4.38	&	24.33	&	0\\
8000	&	5.27	&	28.73	&	0\\
10000	&	6.04	&	32.53	&	0\\
11500	&	6.58	&	35.11	&	0\\
12000	&	6.75	&	35.91	&	0\\
13000	&	7.08	&	37.46	&	1\\
15000	&	7.72	&	40.09	&	1\\[1mm]
\multicolumn{4}{c}{$\epsilon = 0.1$ \ $\Ek = 0.1$ \ $\Pr=0.8$}\\
8000	&	3.17	&	18.19	&	0\\
15000	&	4.81	&	26.2	&	0\\
25000	&	6.5	&	35.04	&	0\\
27000	&	6.8	&	36.59	&	0\\
28700	&	7.04	&	37.86	&	0\\
29300	&	7.13	&	38.3	&	0\\
29400	&	7.13	&	38.37	&	0\\
29500	&	7.15	&	38.44	&	0\\
29650	&	7.17	&	38.55	&	1\\
29700	&	7.18	&	38.58	&	1\\[1mm]
\hline 
\end{tabular}
\label{tab:data}
\end{table}

The clearest way to distinguish those flows where an eye forms 
from those in which it does not is to examine the spatial distribution 
of the azimuthal vorticity, shown in Fig.~\ref{fig10}. 
In cases with a strong eye (Fig.~\ref{fig10}d,h), the 
distribution of $\omega_\phi/r$ is similar to that in Fig.~\ref{fig7}a,
with regions of strong negative vorticity in the boundary layer and
eyewall, and a patch of intense positive vorticity below the eyewall.  
The negative eyewall vorticity has its origins in the boundary layer, with 
the moderately large Reynolds number allowing this vorticity to be swept 
up from the boundary layer into the eyewall. 
Note, in particular, that the magnitude of vorticity in the eyewall is similar to that in the boundary 
layer.

Turning to Fig.~\ref{fig10}b,c,f,g, which are somewhat 
marginal, in the sense that the eye 
is small, we see that overall flow pattern is similar, but that the 
negative eyewall vorticity is now relatively weak, and in particular 
significantly weaker than that in the boundary layer.
It seems likely that the eyewall vorticity is relatively 
weak because advection has to compete with cross-stream diffusion as the 
poloidal flow tries to sweep the boundary layer vorticity up in the eyewall. 
In short, much of the boundary layer vorticity is lost to the surrounding 
fluid by diffusion before the fluid reaches the eyewall. Since the eye 
acquires its vorticity from the eyewall via cross-steam diffusion, the relative 
weakness of the eyewall vorticity explains the weakness of the resulting
eye.

In Fig.~\ref{fig10}a,e, where no eye forms, 
no region of intense negative vorticity 
forms as the poloidal flow turns and $\uu_r$ changes sign. Since there is 
still a reservoir of negative vorticity in the boundary layer, and the basic shape 
of the primary vortex is unchanged, we conclude that the Reynolds number 
is now too low for the flow to effectively sweep the boundary-layer 
vorticity up into the fluid above.

The comparison of Fig.~\ref{fig10}b and Fig.~\ref{fig10}f reveals of
decreasing the Prandtl number yields stronger inertial effects
and thus allows to sustain an eye at lower Rayleigh numbers.

\subsection{A criterion for eye formation}
\label{crit}
The mechanism introduced above works only if the Reynolds number, $\Ree$,
is sufficiently large, so  
that the flow can lift the vorticity out of the boundary layer and into the 
eyewall before is disperses though viscous diffusion. This suggests that there 
is a threshold value in $\Ree$ that must be met in order for an eyewall to form. 
We take this Reynolds number to be based on the maximum value of the inward 
radial velocity, $| \uu_r |_{\rm max}$, at a radial location just 
outside the region containing the eye and eyewall. We (somewhat arbitrarily) 
choose the radial location at which $| \uu_r |_{\rm max}$ is
evaluated to be $r=H$, which does in fact lie just outside the eyewall.  
Thus $\Ree$ is defined as
\begin{equation}
\Ree=\frac{| \uu_r |_{\rm max} H}{\nu}\,,
\label{Re}
\end{equation}
and is chosen to be larger than unity, though not so large that the flow becomes 
unsteady. 

The logic behind this specific definition of $\Ree$ is that the poloidal
velocity field in the vicinity of the eyewall is dominated, via the
Biot-Savart law, by the flux of vorticity up through the eyewall. This, in
turn, is related to the flux of negative azimuthal vorticity in the
boundary layer just outside the eyewall region. However, evaluating this
flux by integrating $|\omega_\phi|$  up through this boundary layer from   to the point
where $|\omega_\phi|=0$ simply gives $| \uu_r |_{\rm max} \, .$ 
We conclude, therefore, that $| \uu_r |_{\rm max}$  is the 
characteristic poloidal velocity in the vicinity of the eyewall, and that 
$\Ree$
is therefore a suitable measure of the ratio of advection to diffusion of
azimuthal vorticity in this region.

\begin{figure}
\centerline{
\includegraphics[width=0.9\textwidth]{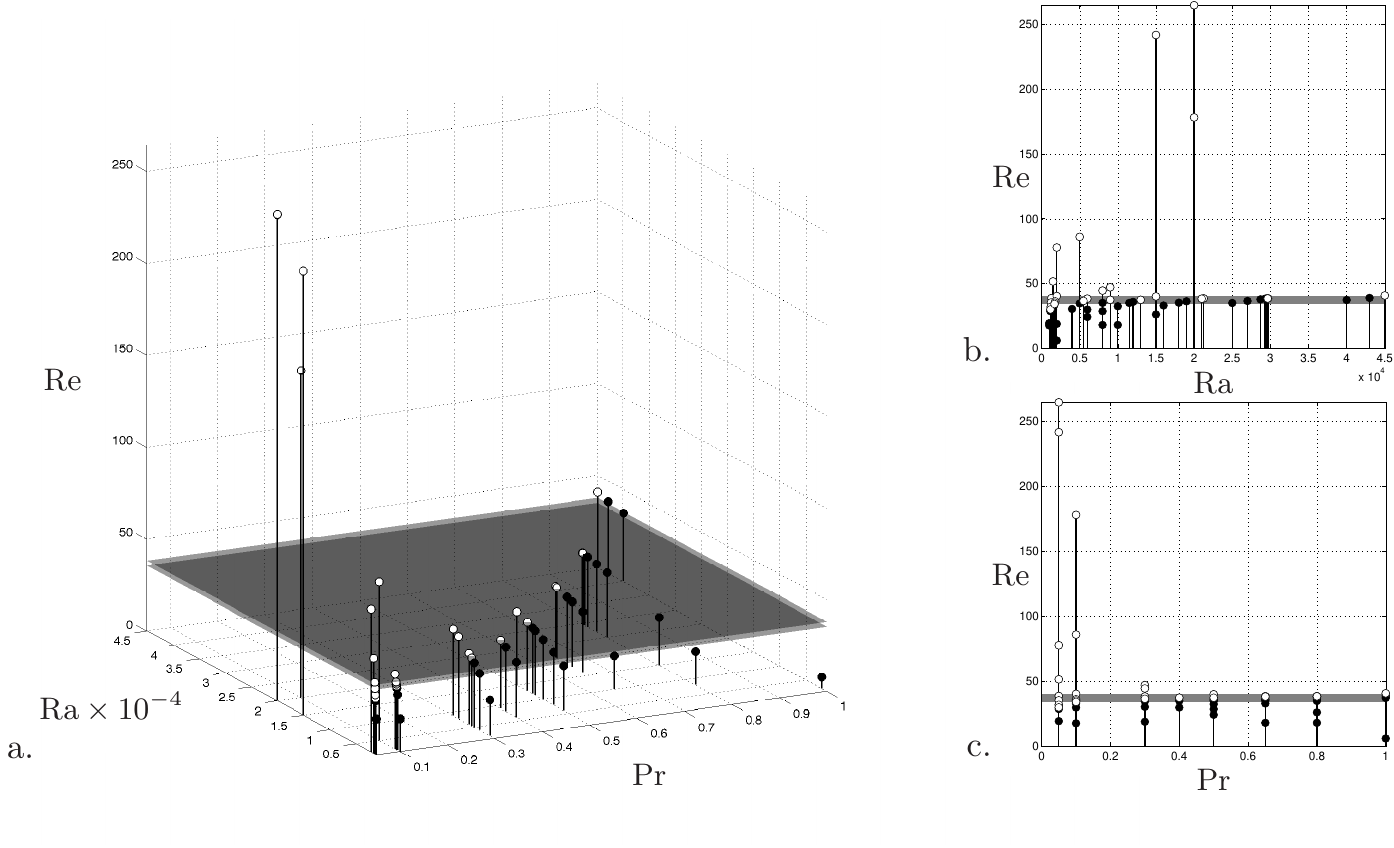}
}
\caption{$\Ree$, as defined by (\ref{Re}), as a function of (a) $\Ray$ and 
$\Pr$, (b) $\Ray$ and (c) $\Pr$. Open symbols represent cases where an eye 
forms, and closed symbols those where there is no eye.}
\label{fig12}
\end{figure}

Note that, since the buoyancy and Coriolis forces are negligible in 
the vicinity of the eye and eyewall, the criterion for a transition from 
eye formation to no eye cannot be controlled explicitly by $\Ek$ or $\Ray$, but 
rather must depend on $\Ree$ only. So, to study the transition in the numerical
experiments, we selected values of $\Ek$, $\Pra$ and $\Ray$ that yield steady flows with
moderately large Reynolds numbers, as defined by \eqref{Re}, and with $\Ro$ somewhat 
larger than unity in the vicinity of the axis, i.e. those conditions which 
are conducive to eye formation.
The values of $\Ree$ obtained from our numerical experiments are
reported in Figure.~\ref{fig12} and some are listed in Table~1.

A reasonably large Reynolds number also turns out to be crucial 
to eye formation. There is indeed a critical value of $\Ree$ below 
which an eye cannot form. Fig.~\ref{fig12}a shows $\Ree$ as a function of $\Ray$ and 
$\Pr$, with the open symbols representing cases where an eye forms, 
and closed symbols those where there is no eye. There is clear 
evidence that the two classes of flow are separated by a plateau in
$\Ree$. 
This is confirmed by the two-dimensional plots of $\Ree$ versus $\Ray$ and
$\Ree$ versus $\Pr$ given in panels (b) and (c). For the model and
parameter regime considered here, the transition occurs at
$\Ree \simeq 37$, as indicated by the horizontal grey surface in Fig.~\ref{fig12}a 
and the corresponding lines in Fig.~\ref{fig12}b,c. 

\section{Conclusions}

We considered axisymmetric steady Boussinesq convection. In the vertical
plane the primary vortex has a clockwise motion, and so has 
positive azimuthal vorticity. In the elongated and rotating domain
considered here, the flow is characterised by a strong swirl as it
approaches the axis.
We have shown that, in this configuration, for sufficiently vigorous
flows, an eye can form. Its motion is 
anticlockwise in the vertical plane, and so the eye is associated 
with negative azimuthal vorticity.
The region that separates the eye from the primary vortex, usually called 
the eyewall,  is characterised by high levels negative azimuthal 
vorticity.
We have shown that it is not generated by 
buoyancy, since such forces are locally too weak. Nor does it arise from 
so-called vortex tilting, despite the local dominance of this process, because 
vortex tilting cannot produce any net azimuthal vorticity. 
We have shown that this thin annular region is filled with 
intense negative azimuthal vorticity, vorticity that has been stripped off 
the lower boundary layer. 
So that the eye 
acquires its vorticity from the surrounding fluid by cross-stream diffusion, 
and this observation holds the key to eye formation in our simple system.

\section*{Acknowledgments}
This work was initiated when
PAD visited the ENS in Paris for one month in May 2015. 
The authors are gratefull to the ENS for support. 
The simulations were performed using HPC resources
from GENCI-IDRIS (Grants 2015-100584 and 2016-100610).

\bibliographystyle{jfm}

\bibliography{ODD_Arxiv}

\providecommand{\noopsort}[1]{}
\begin{thebibliography}{11}
\expandafter\ifx\csname natexlab\endcsname\relax\def\natexlab#1{#1}\fi
\def\au#1{#1} \def\ed#1{#1} \def\yr#1{#1}\def\at#1{#1}\def\jt#1{\textit{#1}}
  \def\bt#1{#1}\def\bvol#1{\textbf{#1}} \def\vol#1{#1} \def\pg#1{#1}
  \def\publ#1{#1}\def\arxiv#1{#1}\def\org#1{#1}\def\st#1{\textit{#1}}

\bibitem[{Batchelor}(1956)]{Batchelor}
{\sc \au{{Batchelor}, G.~K.}} \yr{1956}  \at{{On steady laminar flow with
  closed streamlines at large Reynolds number}}.  \jt{Journal of Fluid
  Mechanics}  \bvol{1},  \pg{177}.

\bibitem[Chandrasekhar(1981)]{Chandra}
{\sc \au{Chandrasekhar, S.}} \yr{1981} {\em Hydrodynamic and hydromagnetic
  stability\/}.  \publ{New York: Dover Publications}, first printed by
  Clarendon Press, 1961.

\bibitem[Davidson(2013)]{Davidson}
{\sc \au{Davidson, P.~A.}} \yr{2013} {\em Turbulence in Rotating, Stratified
  and Electrically Conducting Fluids\/}.  \publ{Cambridge Univ. Press}.

\bibitem[{Drazin}(2002)]{Drazin}
{\sc \au{{Drazin}, P.~G.}} \yr{2002} {\em {Introduction to Hydrodynamic
  Stability}\/}.  \publ{Cambridge Univ. Press}.

\bibitem[{Frank}(1977)]{Frank77}
{\sc \au{{Frank}, W.~M.}} \yr{1977}  \at{{The Structure and Energetics of the
  Tropical Cyclone I. Storm Structure}}.  \jt{Monthly Weather Review}
  \bvol{105},  \pg{1119}.

\bibitem[Guervilly {\em et~al.\/}(2014)Guervilly, Hughes \&
  Jones]{Guervilly2014}
{\sc \au{Guervilly, C.}, \au{Hughes, D.~W.} \& \au{Jones, C.~A.}} \yr{2014}
  \at{Large-scale vortices in rapidly rotating {R}ayleigh{-}{B}\'enard
  convection}.  \jt{Journal of Fluid Mechanics}  \bvol{758},  \pg{407--435}.

\bibitem[Lugt(1983)]{Lugt}
{\sc \au{Lugt, H.~J.}} \yr{1983} {\em Vortex flow in nature and technology\/}.
  \publ{Wiley}.

\bibitem[Pearce(2005a)]{Pearce}
{\sc \au{Pearce, R.}} \yr{2005a}  \at{Why must hurricanes have eyes?}
  \jt{Weather}  \bvol{60}~(1),  \pg{19--24}.

\bibitem[Pearce(2005b)]{Pearce2}
{\sc \au{Pearce, R.}} \yr{2005b}  \at{Comments on ``{W}hy must hurricanes have
  eyes?'' revisited}.  \jt{Weather}  \bvol{60}~(11),  \pg{329--330}.

\bibitem[Rasmussen \& Turner(2003)]{PolarLows}
{\sc \au{Rasmussen, E.~A.} \& \au{Turner, J.}} \yr{2003} {\em Polar Lows,
  MesoscaleWeather Systems in the Polar Regions\/}.  \publ{Cambridge Univ.
  Press}.

\bibitem[Smith(2005)]{Smith}
{\sc \au{Smith, R.~K.}} \yr{2005}  \at{``{W}hy must hurricanes have eyes?''
  revisited}.  \jt{Weather}  \bvol{60}~(11),  \pg{326--328}.

\end{thebibliography}

\end{document}